\documentclass[10pt,twocolumn,superscriptaddress,floatfix,showpacs,nobalancelastpage,amsmath,longbibliography]{revtex4-1}

\usepackage{array,longtable}
\newcolumntype{L}{>{\tiny $}p{0.33\columnwidth}<{$}}
\newcolumntype{M}{>{\scriptsize $}p{0.33\columnwidth}<{$}}
\newcolumntype{N}{>{\scriptsize $}p{0.43\columnwidth}<{$}}
\usepackage{amsmath}
\usepackage{amssymb}
\usepackage{amsfonts}
\usepackage{graphicx}
\usepackage{hyperref}

\usepackage{float}
\usepackage{graphics}

\allowdisplaybreaks[1]

\newif\ifhyper
\hypertrue
\ifhyper
\hypersetup{
   citecolor = {green},
   colorlinks = {true}, 
   urlcolor = {blue} 
}

\newcommand{\dg}{^\dagger}

\DeclareMathOperator{\Tr}{Tr}

\begin{document}


\title{
Entanglement properties of the two-dimensional SU(3) AKLT state
}


%
\author{Olivier Gauth\'e}
\affiliation{Laboratoire de Physique Theorique, C.N.R.S. and Universite de Toulouse, 31062 Toulouse, France}

\author{Didier Poilblanc}

\affiliation{Laboratoire de Physique Theorique, C.N.R.S. and Universite de Toulouse, 31062 Toulouse, France}


\date{\today}


\begin{abstract}  
Two-dimensional (spin-$2$) Affleck-Kennedy-Lieb-Tasaki (AKLT) type valence bond solids on the square lattice  are known to be 
symmetry protected topological (SPT) gapped spin liquids [Shintaro Takayoshi, Pierre Pujol, and Akihiro Tanaka Phys. Rev. B {\bf 94}, 235159 (2016)]. 
Using the projected entangled pair state (PEPS) framework, we extend the construction of the AKLT state to the case of $SU(3)$, relevant for cold atom systems. The entanglement spectrum is shown to be described by an alternating  $SU(3)$ chain 
of   ``quarks" and ``antiquarks", subject to exponentially decaying (with distance) Heisenberg interactions, in close similarity with its $SU(2)$ analog. We discuss the SPT feature of the state. 

\end{abstract}


\pacs{1}
\maketitle

{\it Introduction.}
Recent years have seen growing theoretical interest in systems exhibiting $SU(N)$ symmetry
with the concomitant development of experimental research in condensed matter and atomic physics.
On one hand, many electronic materials possess degenerate low-energy atomic orbitals~\cite{brink_frustration_2011}: (approximate) $SU(N)$ symmetry emerging from this orbital degeneracy leads to interesting physics and is an active field of studies~\cite{Corboz2012,li_su4_1998}.
On the other hand, ultra-cold atom experiments started a new era to design systems with exact $SU(N)$ symmetry~\cite{Bloch2008,Cazalilla2009,Gorshkov2010}.
 Loading these systems on optical lattices, simple lattice models can now be studied in ultra-cold atom ``simulators",
although cooling to low-enough temperature could be a challenge. For example, the fermionic isotopes of alkaline-earth and related elements possess $SU(N)$ symmetry without fine-tuning and can be used to realize various one-dimensional (1D) symmetry-protected topological (SPT) phases in a systematic manner~\cite{Capponi2016,Beverland2016}. These experiments give new perspectives to former theoretical studies of large-$N$
approaches for 1D~\cite{Affleck1985} 
or two dimensional (2D) frustrated quantum magnets~\cite{Read1989} and motivate an increasing number of new studies~\cite{Paramekanti2007,Greiter2007,Nataf2014, Morimoto2014, Capponi2016,Wan2017}.

\begin{figure}
	\centering
		\includegraphics[width=0.3\textwidth]{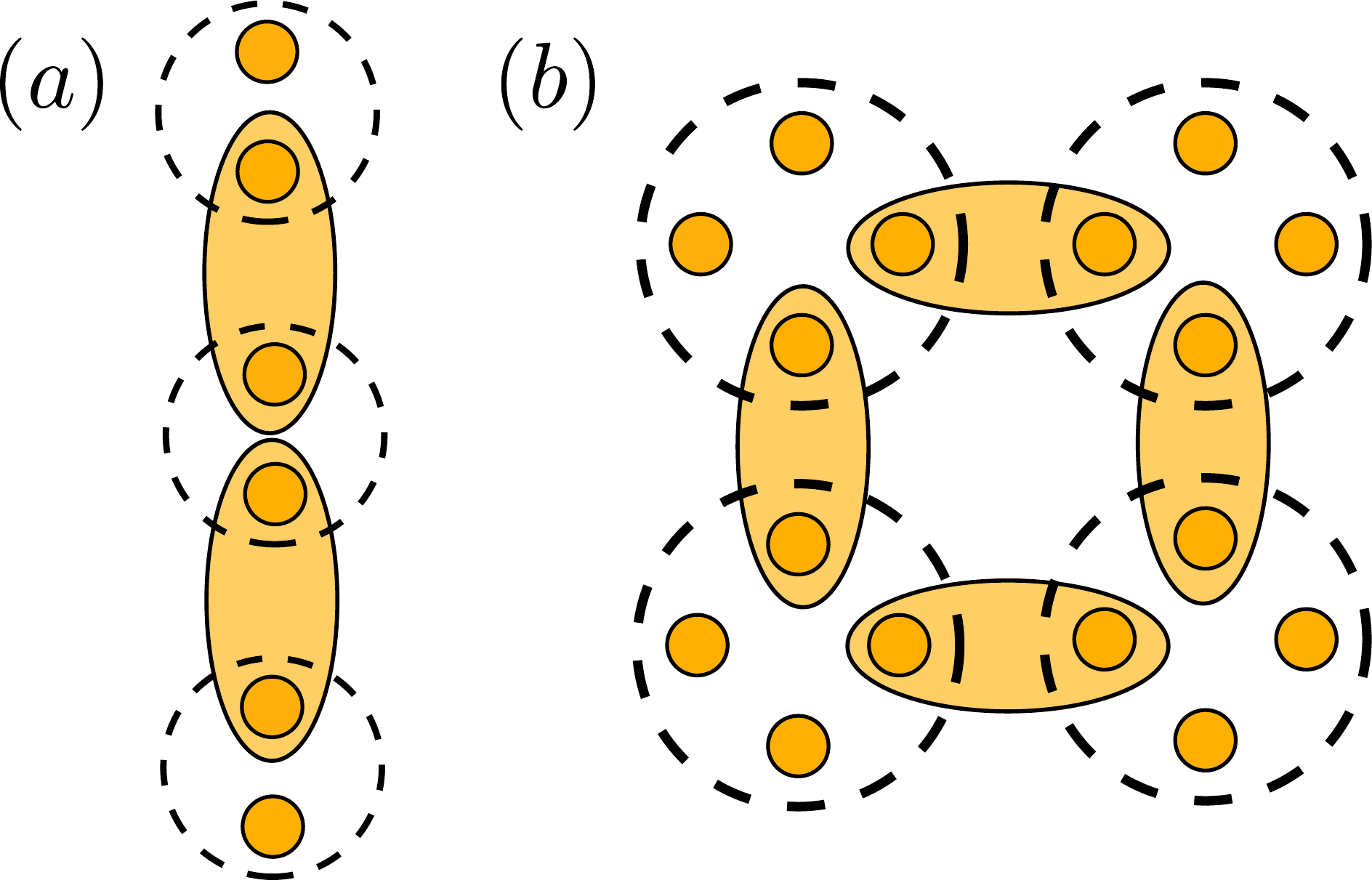}
		\caption{\footnotesize{(Color online)  The SU(2) spin-1 and spin-2 AKLT spin liquids in 1D (a) and 2D (b). Virtual spin-1/2 (orange circles) are entangled into singlets (ellipses). Dashed circles represent projectors on the largest spin irrep.
			}}
		\label{fig:aklt}	
\end{figure}

In condensed matter, electronic or spin systems with spin-$SU(2)$ symmetry are ubiquitous.
In recent years, spin liquids (SL) emerged as a new class of systems defined by the
absence of symmetry breaking, neither lattice nor spin. Soon, Wen introduced the notion of 
topological order (TO) defining a vast category of SL~\cite{Wen1990,Wen2013}. Beyond the conventional 
Ginzburg-Landau paradigm of spontaneous symmetry breaking,
topological spin liquids are characterized by long-range entanglement. 
However, spin liquids can also be short-range entangled, such as e.g. the Haldane chain~\cite{Haldane1983}, the Affleck, Kennedy, Lieb and Tasaki (AKLT) states in 
1D~\cite{Affleck1987a} or 2D~\cite{Affleck1988,Cirac2011}, or the \hbox{spin-$1$} 2D paramagnet~\cite{Jian2016}. The AKLT spin-1 chain was originally defined by i) attaching two spins $1/2$ on every lattice site, ii)
entangling all pairs of spins on the bonds into singlets and iii) projecting pair of spins on every site onto physical spin-$1$.  A pictorial representation of the 1D spin-1 AKLT state is shown in
Fig.~\ref{fig:aklt}(a). The later has a simple parent Hamiltonian defined as a sum of non-commuting projectors on the spin-2 subspace of the nearest neighbor (NN) bonds, i.e. $H_{\rm SU(2)}^{1D}=\sum_i {\cal P}_{i,i+1}^{S=2}=\frac{1}{2}\sum_i({\bf S}_i\cdot{\bf S}_{i+1}+\frac{1}{3}({\bf S}_i\cdot{\bf S}_{i+1})^2+2/3)$, as it can be checked easily that $H_{\rm SU(2)}^{1D}$ is positive definite and annihilates the AKLT state of Fig.~\ref{fig:aklt}(a). Also, it was shown that its fractional spin-$\frac{1}{2}$ edge excitations
(as also in the Haldane chain) are protected by a symmetry~\cite{Pollmann2012}, defining a particular SPT class~\cite{Chen2012,Gu2014,Senthil2015}. 
The AKLT construction can be straightforwardly extended to 2D lattices. On the square lattice (or any 2D lattice of coordination $z=4$), one attaches four virtual spin-$1/2$ on each site, and then projects them onto the most symmetric (i.e. spin-2) irreducible representation (irrep), as shown in Fig.~\ref{fig:aklt}(b). Again, the parent Hamiltonian takes the simple form of a sum of projectors over all NN bonds $\langle i,j\rangle$, $H_{\rm SU(2)}^{2D}=\sum_{\langle i,j\rangle}{\cal P}_{i,j}^{S=4}$. In 2D, the family of AKLT states are protected by $SU(2)$ spin-rotations and one-site translation symmetries~\cite{Takayoshi2016}, a direct consequence of the Lieb-Schultz-Mattis~\cite{Lieb1961} theorem.

Tensor network techniques offer a new versatile method to construct simple 1D or 2D paradigmatic wave functions such as AKLT states or resonating valence bond SL~\cite{Anderson1973,Schuch2012,Poilblanc2012}. 
Matrix product states (MPS) and projected entangled pair states 
(PEPS)~\cite{Cirac2009b,Cirac2012a,Orus2013,Schuch2013b,Orus2014} are simple 
1D and 2D ans\"atze, constructed from a single site matrix or tensor, respectively.
$SU(2)$-symmetric PEPS can be classified according to their lattice symmetries~\cite{Poilblanc2016}, 
allowing to construct systems with tunable symmetries and exotic properties. 
In addition, the PEPS framework enables to compute entanglement properties\cite{Cirac2011, Lou2011}  
-- entanglement spectrum (ES), entanglement Hamiltonian (EH), etc... -- in a very efficient way. 
It turns out that the 1D or 2D $SU(2)$ AKLT states have extremely simple representations in terms of MPS~\cite{Pollmann2012} and PEPS~\cite{Cirac2011}, respectively, which make the analysis of their bulk and edge properties accurately computable. 

Although AKLT parent Hamiltonians are fine-tuned, the AKLT states provide in fact simple paradigms for
the simplest (non-topological) gapped spin liquid phases, which can occupy a rather extended region
in the parameter space of realistic Hamiltonians. 
For example, the 1D spin-1 AKLT state corresponds to a special point of the well-known extended Haldane phase describing
several experimental spin-1 chains. Since localized $SU(N)$ spin systems can now be realized on optical 1D and 2D lattices,
$SU(N)$ AKLT states are expected to describe generic spin liquid phases in such systems and are therefore of high interest. 
In the case of a SPT phase, the edge modes of the AKLT wave function will also be generic of the whole phase, 
being protected by symmetry. 
In this rapid communication, we extend the 2D AKLT state to the case of  
 $SU(3)$ symmetry. We show that it can be represented 
as a simple tensor network, allowing for extensive studies. We explore its bulk properties on an infinite cylinder, using transfer matrix methods.
The edge physics is investigated by computing the entanglement spectrum and the related entanglement Hamiltonian. We show that the latter can be very well approximated 
by a simple $SU(3)$ Heisenberg Hamiltonian with exponentially decaying interactions.

\begin{figure}
	\centering
		\includegraphics[width=0.5\textwidth]{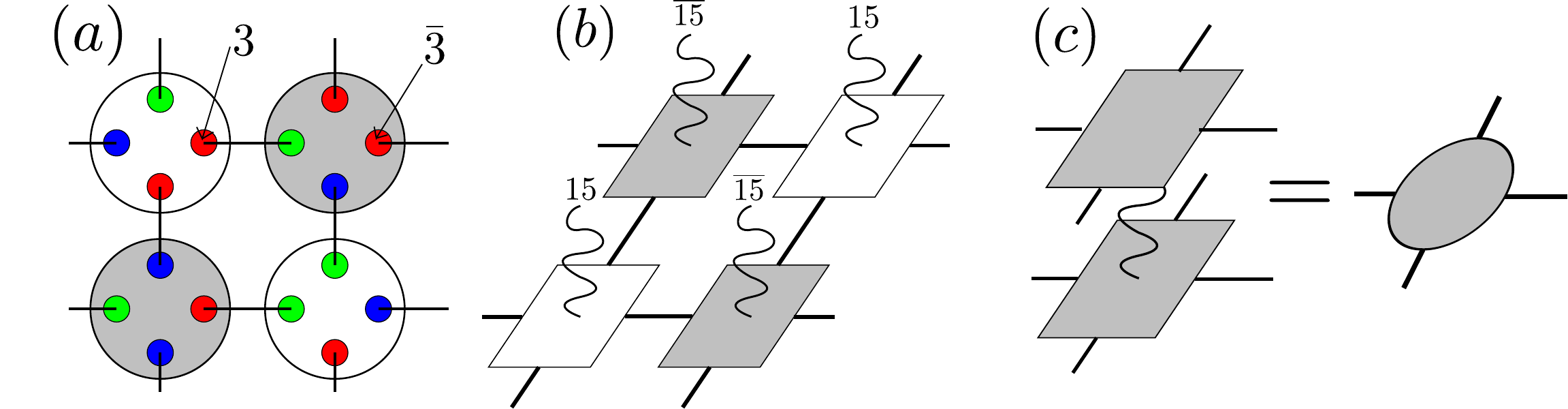}
		\caption{\footnotesize{(Color online)  (a,b) The AKLT $SU(3)$ wave function is defined similarly to the $SU(2)$ case: four virtual states in the fundamental (anti-fundamental) irrep of $SU(3)$ of dimension $D=3$, are attached on even (odd) sites
and projected onto the fully symmetric $ \textbf{15}$ ($\overline{\textbf{15}}$) irrep. 
Virtual states of all neighboring sites are projected on $SU(3)$ singlets
to form a tensor network. (c) By contracting two identical site tensors on their physical indices one gets a new tensor $\mathbb{E}$ of dimension $D^2=9$. 
		}}
		\label{fig:pannel}	
\end{figure}

{\it $SU(3)$ AKLT wavefunction.}
We now extend the recipe for the construction of $SU(2)$ AKLT states to $SU(3)$, in a straightforward way. 
In that case, we use standard Young tableau notations to label the
$SU(3)$ irreps or ``spins" (also denoted by their dimension in bold).  First, in order to realize $SU(3)$ singlets on all NN bonds of the square lattice, four ``quarks" in the fundamental $[1]={\bf 3}$ irrep (``antiquarks" in the anti-fundamental $[1,1]={\overline{\bf 3}}$ irrep) are attached on each even (odd) site. This way, neighboring virtual spins on every NN bond belong to $\textbf{3}$ and $ \overline{\textbf{3}}$ irreps and can then be projected onto $SU(3)$
$[1,1,1]={\bf 1}$ singlets. Then, in order to entangle this simple product of singlets,
one projects the group of four quarks on each even (odd) site onto the most symmetric $[4]=\textbf{15}$
($[4,4]= \overline{\textbf{15}}$) irrep corresponding to the actual physical degrees of freedom, as seen in figure \ref{fig:pannel}(a). 
Note that the assignment as fundamental or anti-fundamental is arbitrary, the same tensor being placed on every site. 
As for $SU(2)$, a simple parent Hamiltonian can be build from bond projectors on the largest, most-symmetric $[8,4]$ (self-conjugate) irrep obtainable from the tensor-product $\textbf{15}\otimes \overline{\textbf{15}}$, 
\begin{equation}
\mathcal{H}_{SU(3)}^{2D}=\sum_{\langle i,j\rangle}{\cal P}_{i,j}^{[8,4]}\, ,
\end{equation}
where the sum runs over all NN bonds. 

{\it Description of the PEPS formalism.} For simplicity, let us first start with a periodic ($L$-site) 1D chain with $d$ on-site physical degrees 
of freedom labeled by $\alpha$ (e.g. the components of the physical spin).
By definition, the amplitudes $c_{\alpha_1\alpha_2\cdots\alpha_L}$ of  a (translational-invariant) MPS of virtual dimension $D$ 
are given solely
in terms of $d$ $D\times D$ matrices $A^{\alpha}$ as $c_{\alpha_1\alpha_2\cdots\alpha_L}=\Tr \{ A^{\alpha_1} 
A^{\alpha_2}\cdots A^{\alpha_L} \}$. 
It is easy to see that the 1D SU(2) AKLT state of Fig.~\ref{fig:aklt}(a) is in fact a MPS defined from a set of three
$2\times 2$ matrices labelled by the physical spin (i.e. $d=3$ and $D=2$). 
This construction can easily be generalized in 2D by replacing the $d$ matrices by $d$ rank-$z$ 
tensors, where $z$ is the lattice coordination number ($z=4$ in our case). 
The amplitudes of the PEPS are then obtained from the tensor network defined by attaching a tensor on each lattice site and by contracting the site tensors over the virtual indices~\cite{Cirac2009b,Cirac2012a,Orus2013,Schuch2013b,Orus2014}.
The $S=2$ AKLT state of Fig.~\ref{fig:aklt}(b) can then be viewed as a simple PEPS with $D=2$ virtual degrees of freedom (corresponding to the attached virtual spin-1/2) and $d=2S+1=5$ physical spin components~\cite{Cirac2011}.
Similarly, the $SU(3)$ AKLT state of Fig.~\ref{fig:pannel}(b) can be interpreted as a PEPS of virtual dimension $D=3$ 
(for the three colors of the quarks) and $d=15$ physical dimension, as depicted in Fig.~\ref{fig:pannel}(b).

\begin{figure}
	\centering
		\includegraphics[width=0.4\textwidth]{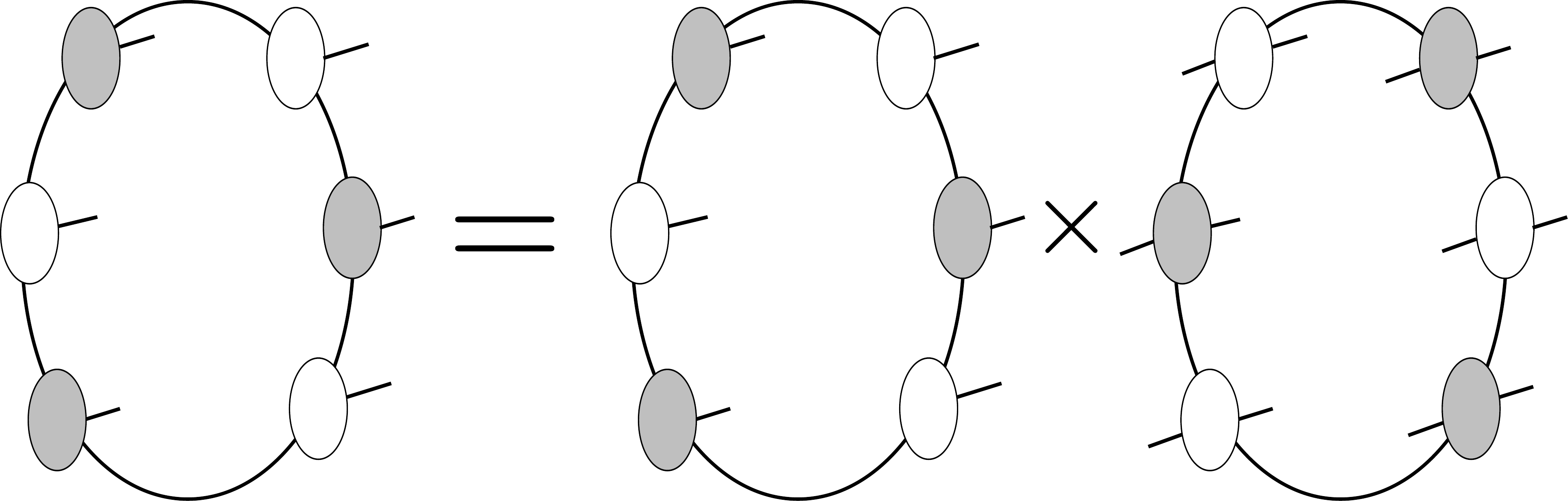}
		\caption{\footnotesize{The fixed-point boundary state is defined as the leading eigenvector of the transfer matrix. The latter is defined by contracting the local $\mathbb{E}$ tensor along a circle, leaving the left and right legs open.
		}}
		\label{fig:cylinder}	
\end{figure}

In practice, one needs to compute the PEPS wave function norm $\langle\Psi |\Psi\rangle$ or expectation values $\langle\Psi |O|\Psi\rangle$ of local operators $O$. For such purpose, one first defines a two-layer tensor network, each layer representing the ket and bra 
wave functions. By contracting two identical tensors on their physical indices one gets a new tensor $\mathbb{E}$ of dimension $D^2=9$, as shown in figure \ref{fig:pannel}(c). This way, the physical index disappears and its large dimension ($15$) is irrelevant for computations. We form an infinite cylinder by imposing periodic boundary conditions in one direction with circumference $N_v$.
Each row of the cylinder can then be seen as a transfer matrix, propagating states from the left to the right. This matrix acts on boundary states expressed in terms of virtual variables of the tensor network as shown in figure \ref{fig:cylinder}. To construct the fixed point boundary state of size $(D^2)^{N_v}$, one uses iterated powers / Lanczos algorithm to converge to the leading eigenvector / leading eigenvalues of the transfer matrix. Note that since the latter is a symmetric matrix, the left and right boundary states are identical.

\begin{figure}[ht]
	\centering
		\includegraphics[width=0.50\textwidth]{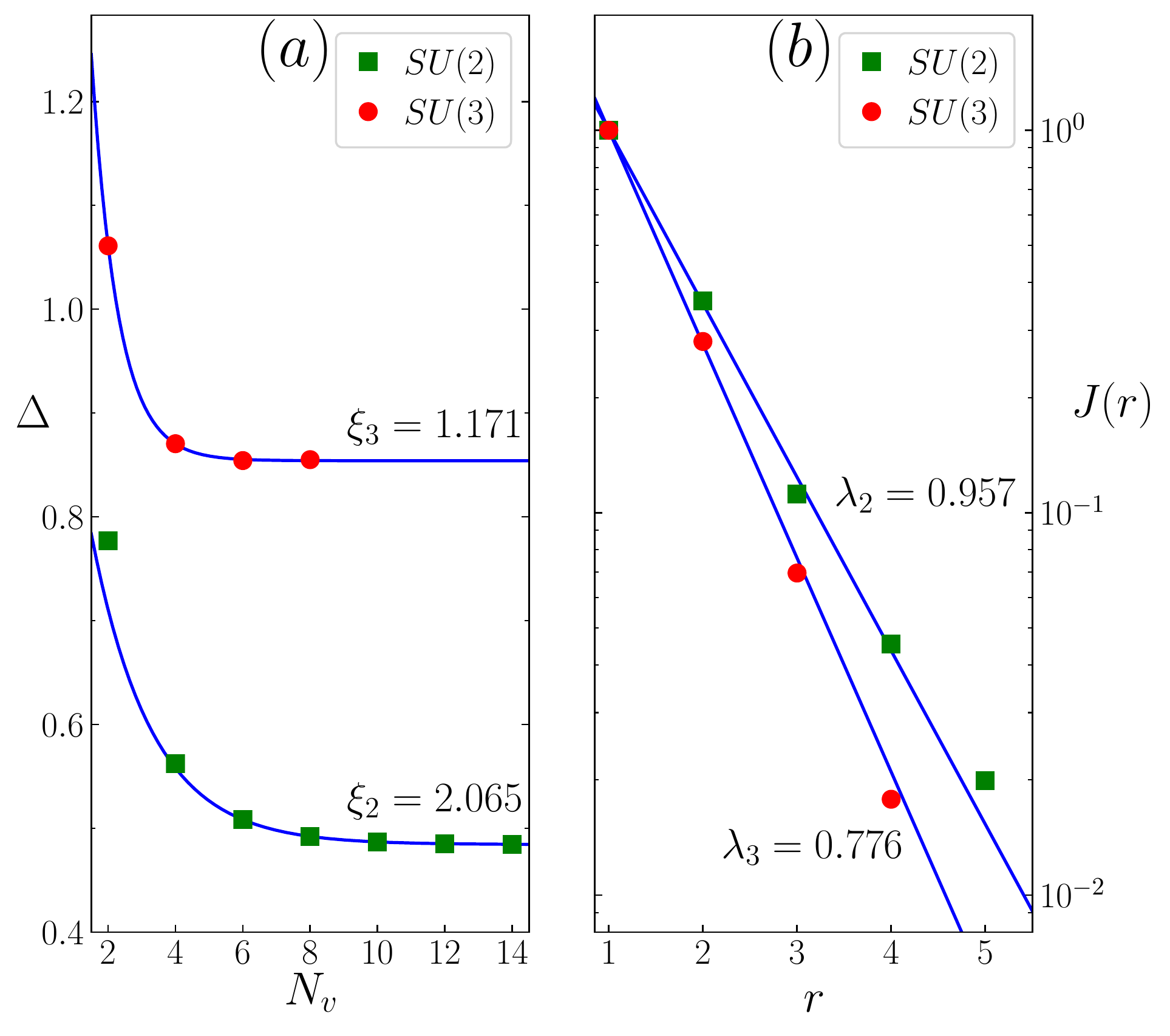}
		\caption{\footnotesize{(Color online) (a) Bulk gap of an infinite AKLT SU(N) cylinder vs circumference $N_v$. The extrapolated $N_v\rightarrow\infty$ values of $\xi = 1/\Delta$ are shown on the plot. (b) Coefficients of the effective entanglement Hamiltonian (decomposed in term of Heisenberg-like operators) for $SU(2)$ and $SU(3)$ AKLT wavefunctions vs site separation (semi-log plot). Straight lines are fits according to an exponential behavior $J(r)=J_0 \exp{(-r/\lambda)}$. Data for $SU(2)$ are taken from reference~\cite{Cirac2011}. }}
		\label{fig:gap-weights}
\end{figure}

{\it Bulk properties.} The gap $\Delta$ in the bulk can easily be computed from the two largest eigenvalues of the transfer matrix, $\Delta = \ln{(E_1/E_2)}$, with $E_1 > E_2$, the correlation length $\xi$ being defined as the inverse of the gap. We have computed $\Delta$ for cylinders of perimeter $N_v = 2, 4, 6, 8$ and extrapolated the result in the limit $N_v \rightarrow \infty$, as shown in figure \ref{fig:gap-weights}(a). We find that the extrapolation of $\xi$ for the $SU(3)$ case is very short ($\xi_3 \simeq 1.2$), even shorter than the $SU(2)$ value ($\xi_2 \simeq 2.1$). Note that the extrapolation is very accurate, the scaling being exponential and the system size being large compared to $\xi$.

\textit{Entanglement Hamiltonian and entanglement spectrum.}
In order to construct the entanglement Hamiltonian (EH), the fixed-point state (see above) is reshaped as a $D^{N_v} \times D^{N_v}$ boundary density matrix $\Sigma_b$, acting on virtual variables. It has previously been shown~\cite{Cirac2011} that this matrix can be mapped onto the reduced density matrix of the half cylinder $\rho$ via an isometry,
$\rho = U\dg (\Sigma_b)^2 U$. The entanglement Hamiltonian $\mathcal{H}$ acting on virtual boundary configurations is defined via
$(\Sigma_b)^2 = \exp(-\mathcal{H})$.
 
\begin{figure}
	\centering
		\includegraphics[width=0.40\textwidth]{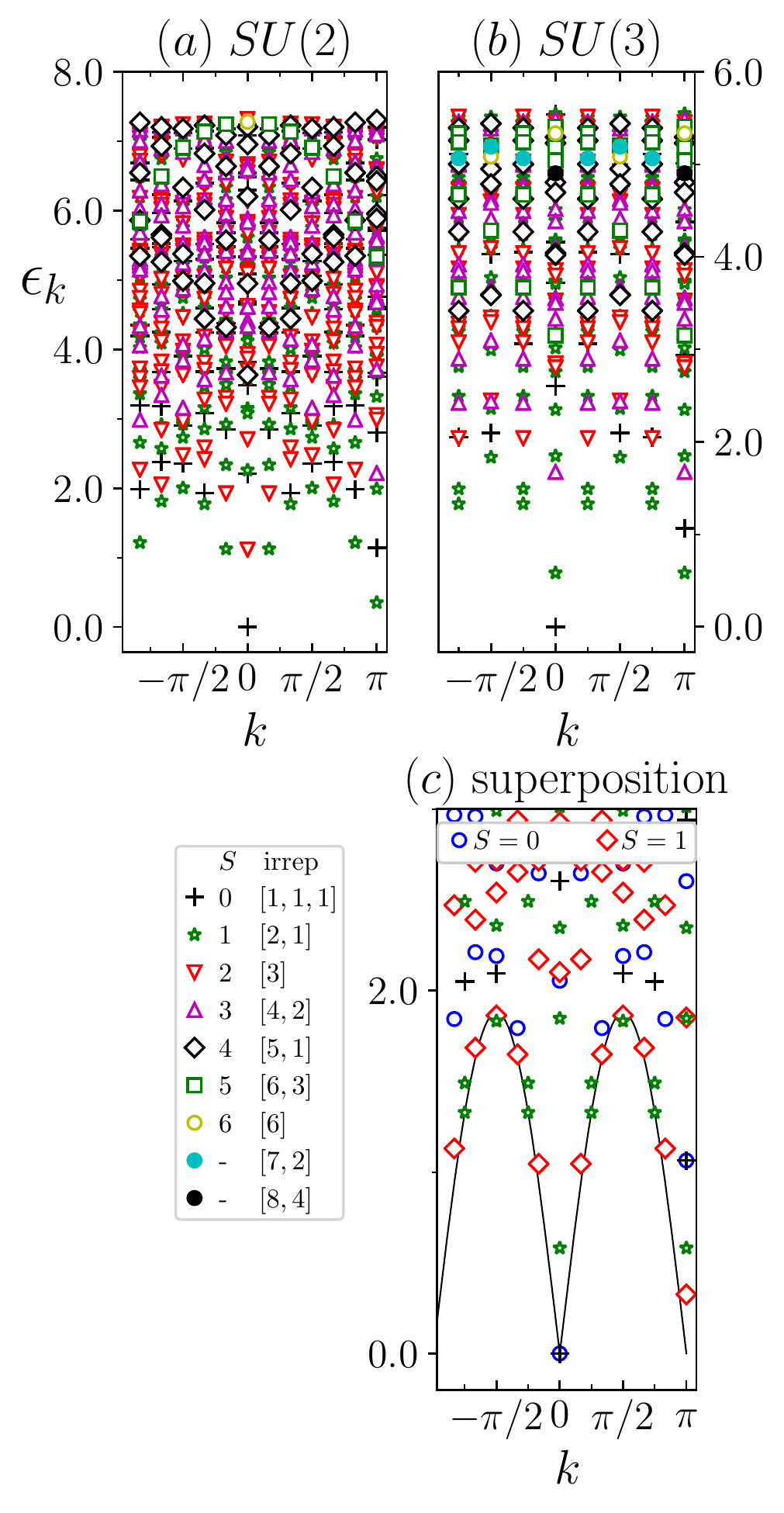}
		\caption{(Color online) Entanglement spectra on infinite cylinders of finite circumference $N_v$. (a) $SU(2)$ AKLT wavefunction computed with $N_v=12$, irreps are indexed by their spin.  
		(b) $SU(3)$ AKLT wavefunction computed with $N_v=8$, irreps are indexed according to their Young tableaux.
		(c) Comparison of the low-energy part of the two spectra superposed on the same graph (only trivial and adjoint irrep are kept, with new symbols for the SU(2) spectrum). The $SU(2)$ spectrum is rescaled to match the first singlet excitation 
		(at $k=\pi$) of the two spectra. Lines are sinusoidal fits of the edge of the 2-spinon continuum.}
		\label{fig:spectra}	
\end{figure}

 The spectrum of $\mathcal{H}$ -- the entanglement spectrum (ES) -- has been 
conjectured by Li and Haldane~\cite{Li2008}, to be in one-to-one correspondence with the physical edge modes of the system.
We compare the ES of $SU(2)$ and $SU(3)$ AKLT wavefunctions in figure \ref{fig:spectra}. We observe they are very much similar at low energy: (i) the ground state is a singlet with momentum $k=0$ (when $N_v = 4n$), (ii) low-energy excitations follow a sinusoidal dispersion typical of the lower edge of a 2-spinon continuum, shown in figure \ref{fig:spectra}(c). This can be explained from the simple (approximate) analytical form of the EH (derived next).

To understand its nature we decompose the EH on the canonical basis of $SU(3)$ operators acting on the virtual degrees of freedom at the boundary. The latter are being defined in a fermionic representation as 
\begin{equation}
S^\alpha_\beta (i) =  \left\{
  \begin{array}{l l}
    c\dg_{\alpha,i} c_{\beta,i} - \delta_{\alpha,\beta}/3  \quad \text{if $i$ is even}\\
   c_{\alpha,i} c\dg_{\beta,i} - \delta_{\alpha,\beta}/3  \quad \text{if $i$ is odd}
  \end{array} \right.
\end{equation}
where $\alpha, \beta$ label the three $SU(3)$ colors. Note that the definition takes into account the anti-fundamental representation on odd sites~\cite{Affleck1985}, which in the fermion language is obtained via a particle-hole transformation.
Since the Hamiltonian is $SU(3)$ invariant, there is a limited number of combination of operators that can appear, in particular no linear term can appear. The only second order $SU(3)$ invariant terms are Heisenberg-like terms, $\textbf{S}_i \cdot \textbf{S}_j = \sum_{\alpha,\beta} S^\alpha_\beta (i) S^\alpha_\beta (j)$. Hence,
\begin{equation}
\mathcal{H} = E_0 + \sum_{i\neq j} J(|i-j|) \,\textbf{S}_ i\cdot \textbf{S}_j + \mathcal{H}_\text{rest}
\label{eq:H_Heisenberg}
\end{equation}
where $E_0 = \Tr(\mathcal{H})$. The higher order terms $\mathcal{H}_\text{rest}$ are corrections of much lower weights --  only $5\%$ (6\%) of the euclidean norm of  $\mathcal{H} - E_0 $ for $N_v = 8$ ($N_v = 6$) -- and are expected to be irrelevant.
We show in figure \ref{fig:gap-weights}(b) that the weights $J(r)$ follow an exponential decay with distance, from with we can extract a typical decay length $\lambda$. By comparing $SU(3)$ and $SU(2)$, we see that $\lambda_3<\lambda_2$, fulfilling the same inequality than the bulk correlation length $\xi_3<\xi_2$. This is in agreement with a general argument based 
on PEPS that the
range $\lambda$ of the EH tracks the bulk correlation length $\xi$~\cite{Cirac2011}.

\textit{Discussion and outlook.}
Interestingly, the EH of the $SU(3)$ AKLT state is adiabatically connected to the nearest neighbor $\textbf{3}-\bar{\textbf{3}}$ Heisenberg chain~\cite{Affleck1985}. The latter can be mapped to a spin-$1$ chain with a purely negative biquadratic coupling and was shown to
exhibit a small spontaneous 
dimerization~\cite{Affleck1987b,Affleck1989,Barber1989,Sorensen1990}. It is however plausible that the extra  $J(2)\sim 0.3\, J(1)$ coupling will close the gap and lead to a gapless spectrum. Indeed, the numerical ES shown in Figs.~\ref{fig:spectra}(b,c) does not show any hint of spontaneous translation symmetry breaking (implying GS two-fold degeneracy in the $N_v\rightarrow\infty$ limit).
The conformal field theory (CFT) description of our EH is an open problem which would require the numerical treatment of very long chains.
Interestingly, the parent 
Hamiltonian~\cite{Tu2014,Bondesan2014} for a
 CFT wave function constructed from the $SU(3)_1$ Wess-Zumino-Witten (WZW) models~\cite{Witten1983} is, once truncated, quite similar to our quasi-local EH, although with a larger ratio  $J(2)/J(1)\simeq 0.56$ and a 3-body term of significant amplitude. Hence a description of the EH in terms of a $SU(3)_1$ WZW theory seems natural and, at least, agrees with our low-energy ES shown in figure \ref{fig:spectra}(c). 
Tu et al.~\cite{Tu2014} report critical properties deviating from the expected behaviors of the $SU(3)_1$ WZW model. We note however that the two
(local) models may sit in different critical phases. 

Another interesting question is the possible correspondence between the ES and the edge physics~\cite{Li2008}.
As for the $SU(2)$ AKLT state, one can construct a local $SU(3)$-invariant parent Hamiltonian or 
``PEPS model"~\cite{Perez-Garcia2007,Yang2014} for which, any region with an open 1D boundary $\partial R$ will have a degenerate manifold of (at most) $D^{|{\partial R}|}$ GS. As for any PEPS models in a trivial (i.e. short-ranged entangled) phase, any Hamiltonian can be realized on the edge~\cite{Yang2014} by slightly perturbing the
(fine-tuned) $SU(3)$ PEPS model. 
However, it is still possible to protect edge properties by
symmetries in the bulk~\cite{Chen2012}. For example, similarly to the $SU(2)$ AKLT model,
$SU(3)$ symmetry and translation invariance rule out a gapped edge which does not break any symmetry~\cite{Lieb1961}. This is in direct correspondence with the properties of the (infinite size) ES discussed above.

Lastly, we comment on the relevance of this work to cold atoms physics. 
Constructing bipartite lattices of localized $SU(3)$ atoms in staggered conjugate irreps 
is possible experimentally although challenging~\cite{Laflamme2016}. It is also
of interest to enforce the same irrep on every site.
For this goal, a different AKLT construction exits, involving virtual states belonging to the 
smallest self-conjugate irrep. For $SU(3)$ it corresponds to the adjoint $[2,1]$ (8-dimensional) 
irrep. The case of $SU(4)$ would be simpler using the self-conjugate 
(antisymmetric) $[1,1]$ (6-dimensional)~\cite{Wan2017} 
irrep for the virtual states. The physical site degrees of freedom on a 2D square lattice correspond to atoms in the $[4,4]$ ($105$-dimensional) irrep of $SU(4)$. 

{\it Acknowledgments.}   
We are grateful to Fr\'ed\'eric Mila, Philippe Lecheminant and Hong-Hao Tu for useful suggestions and pointing out interesting references. DP also thanks Takahiro Morimoto, Masaki Oshikawa, Karlo Penc and Pierre Pujol for enlightening discussions. 
OG thanks Mikl\'os Lajk\'o for help with the definition of $SU(3)$ operators and Xavier Bonnetain for his Young tableaux product calculator.
%





\bibliography{bibliography2} 



\end{document}
